\documentclass[twocolumn,aps,prl,showpacs]{revtex4}

\usepackage{epsfig}

\begin{document}

\title{Novel neutron resonance mode in $d_{x^2-y^2}$ superconductors}
\author{Ilya Eremin$^{1,*}$, Dirk K. Morr$^{1,2}$, Andrey V. Chubukov$^3$,
Karl Bennemann$^1$, and Michael R. Norman$^{4}$ }

\affiliation{$^1$ Institute fur Theoretical Physick, Free
Universit\"{a}t Berlin, D-14195, Berlin, Germany \\ $^2$
Department of Physics, University of Illinois at Chicago, Chicago,
IL 60607\\ $^3$ Department of Physics, University of Wisconsin,
Madison, WI 53706\\ $^4$ Materials Science Division, Argonne
National Laboratory, Argonne, IL 60439}
\date{\today}

\begin{abstract}
We show that a new resonant magnetic excitation at incommensurate
momenta, observed recently by inelastic neutron scattering
experiments on YBa$_2$Cu$_3$O$_{6.85}$ and YBa$_2$Cu$_3$O$_{6.6}$,
is a {\it spin exciton}. Its location in the magnetic Brillouin
zone and its frequency are determined by the momentum dependence
of the particle-hole continuum. We identify several features that
distinguish this novel mode from the previous resonance mode
observed near ${\bf Q}=(\pi,\pi)$, such as its intensity maximum
which occurs in a different part of the magnetic Brillouin zone.
\end{abstract}

\maketitle

While it seems established that the resonance peak is a universal
feature of the high-temperature superconductors \cite{YBCO,Bi,Tl},
its origin, its role for the pairing process, and the effects
arising from its interactions with electrons are still intensively
debated \cite{theory}. The peak's intensity is the highest at
${\bf Q}= (\pi,\pi)$ and in YBa$_2$Cu$_3$O$_{6+x}$ (YBCO), where
it was studied in great detail, its frequency $\Omega_{res}({\bf
Q})$ follows the same doping dependence as $T_c$, with
$\Omega_{res}({\bf Q}) \approx 41$ meV near optimal doping. As one
moves away from ${\bf Q}$, the peak disperses downwards and its
intensity decreases rapidly, vanishing around ${\bf Q}_0 =
(0.8\pi, 0.8 \pi)$.  The  doping dependence of $\Omega_{res} ({\bf
Q})$, the downward dispersion of the resonance, and the fact that
${\bf Q}_0$ coincides with the distance between nodal (diagonal)
points on the Fermi surface are consistent with the theoretical
idea that the resonance peak is a particle-hole bound state below
the spin gap (a spin exciton) \cite{excit} (for a review of other
theoretical scenarios, see Ref.~\cite{other}).

Recent inelastic neutron scattering (INS) experiments on
YBa$_2$Cu$_3$O$_{6.85}$ \cite{pai} in the superconducting (SC)
state detected a new resonant magnetic excitation at
incommensurate momenta, but at frequencies {\it larger} than
$\Omega_{res}({\bf Q})$. A similar result was obtained for
underdoped YBa$_2$Cu$_3$O$_{6.6}$ \cite{Hay04}. This new resonance
mode is particularly pronounced along the diagonal of the magnetic
Brillouin zone (MBZ) at ${\bf q} \lesssim {\bf Q}_0$. It was
suggested \cite{pai} that this new resonance is a particle-hole
bound state with an upward dispersion originating at ${\bf Q}$
(see Fig.~5a in Ref.~\cite{pai}).

In this Letter, we show both analytically and numerically that the
new resonance mode is indeed a spin exciton that emerges below
$T_c$ due to a feedback effect on the collective spin excitations
arising from the opening of the SC gap in a $d_{x^2-y^2}$
superconductor. We demonstrate that the new resonance appears only
at momenta less than ${\bf Q}_0$, and is separated from the
previous resonance by a region near ${\bf Q}_0$ in which no
resonance exists (the ``silent band" of Ref.~\cite{pai}). Thus,
the new resonance does {\it not} form an upward dispersing branch
originating at ${\bf Q}$.  We identify several qualitative
features that distinguish this new resonance (the $Q^*$ mode) from
the old one (the $Q$ mode). In particular, we show that while the
intensity of the $Q$ mode is largest along ${\bf q}=(\pi,\eta
\pi)$ and ${\bf q}=(\eta \pi, \pi)$, the $Q^*$ mode has its
largest intensity along the MBZ diagonal.

We begin by presenting our numerical analysis of the new $Q^*$
mode. Its emergence can be understood within an RPA approach for
which the spin susceptibility is given by
\begin{equation}
\chi({\bf q},\omega)=\frac{\chi_0({\bf q},\omega)}{1-g({\bf q})
\chi_0({\bf q},\omega)} \label{fullchi}
\end{equation}
where $g({\bf q})$ is the fermionic four-point vertex, and
$\chi_0({\bf q},\omega)$ is the free-fermion susceptibility, which
in the SC state is given by the sum of two single bubble diagrams consisting
of either normal or anomalous Greens functions
\cite{excit,com5}. For our numerical calculation of
$\chi_0({\bf q},\omega)$, we used a SC gap with
$d_{x^2-y^2}$ symmetry and a normal state tight binding
dispersion
\begin{equation}
\epsilon_{\bf k}=-2t \left( \cos k_x + \cos k_y \right) - 4 t'
\cos k_x \cos k_y - \mu \label{Disp}
\end{equation}
with $t=250$ meV, $t'/t=-0.4$, and $\mu/t=-1.083$. The Fermi
surface (FS) obtained from Eq.~(\ref{Disp}) is shown in
Fig.~\ref{FS}(a). It describes well the FS measured by
photoemission experiments on
Bi$_2$Sr$_2$CaCu$_2$O$_{8+\delta}$~\cite{ARPES}.

\begin{figure}[h]
\epsfig{file=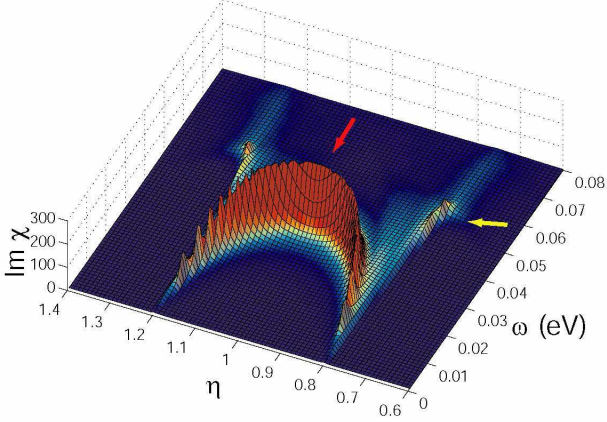,width=8.5cm} \caption{(color) RPA results for
magnetic excitations in a $d_{x^2-y^2}$ superconductor. Im$\chi$
obtained from Eq.~(\ref{fullchi}) as a function of momentum (along
${\bf q}=\eta (\pi,\pi)$) and frequency in the SC state.  We used
$\Delta_{\bf k} = \Delta_0 (\cos k_x - \cos k_y)/2$ with
$\Delta_0=42$ meV, and $g({\bf q})=g_0[1-0.1(\cos q_x +\cos q_y)]$,
with $g_0=0.573$ eV in order to reproduce the correct energy
position of the $Q^*$ mode near $0.8(\pi,\pi)$ and the $Q$ mode at
$(\pi,\pi)$.} \label{intensity}
\end{figure}

\begin{figure}[h]
\epsfig{file=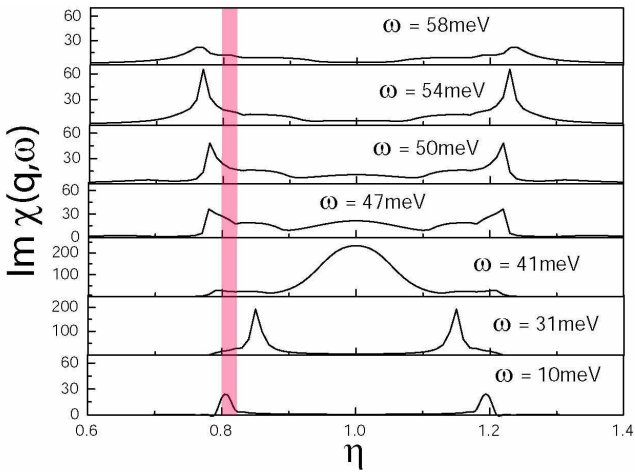,width=8.5cm} \caption{Im $\chi({\bf
q},\omega)$ along ${\bf q}=\eta (\pi,\pi)$ for several frequencies.
The position of the silent band is indicated by the shaded area.
These results reproduce the experimental INS data, see Fig.~2(a) in
Ref.~\cite{pai}. } \label{exper}
\end{figure}

\begin{figure}[h]
\epsfig{file=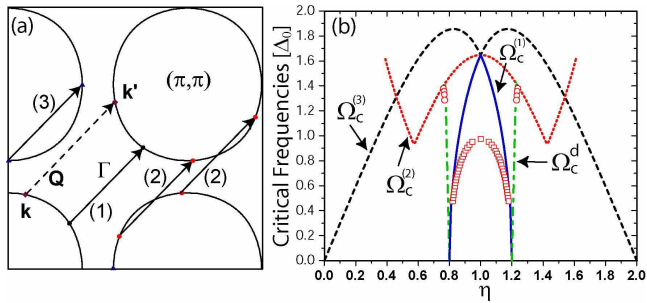,width=8.5cm} \caption{(a) FS of
Eq.~(\ref{Disp}) and magnetic scattering vectors. (b) Momentum
dependence of the critical frequencies $\Omega_{c}^{(i)}$
($i=1,2,3$) and the direct gap $\Omega_{c}^{d}$ along ${\bf q}= \eta
(\pi,\pi)$ (see text). The open squares (circles) represent the
position of the $Q$ ($Q^*$) resonance.} \label{FS}
\end{figure}

Our main results are presented in Figs.~\ref{intensity} and
\ref{exper}, in which we plot Im$\chi ({\bf q}, \omega)$, obtained
from a numerical evaluation of Eq.(\ref{fullchi}), along ${\bf
q}=\eta (\pi,\pi)$ in the SC state.
The intensity plot of Im$\chi ({\bf q}, \omega)$ shown in
Fig.~\ref{intensity} possesses all salient features observed in
the INS experiments. First, we identify a downward dispersion of
the $Q$ mode (indicated by a red arrow). Second, the $Q^*$ mode
(indicated by a yellow arrow) is located at frequencies larger
than the frequency of the $Q$ mode at $(\pi,\pi)$, and is confined
to a small region of momentum space near ${\bf Q}_0$. The momentum
position of the $Q^*$ mode is almost independent of energy. Third,
there exists a region in momentum space around ${\bf Q}_0$ that
separates the $Q$ mode from the $Q^*$ one (the ``silent band" of
Ref.~\cite{pai}). Note, that as ${\bf Q}_0$ is approached from
$(\pi,\pi)$, the $Q$ mode frequency, $\Omega_{res}({\bf q})$, as
well as its intensity, rapidly decreases. In Fig.~\ref{exper} we
present Im$\chi$ along ${\bf q}=\eta (\pi,\pi)$ for several
frequencies. We clearly see that the two modes are separated in
momentum and frequency space. The shaded area in Fig.~\ref{exper}
represents the silent band, in which Im$\chi$ is strongly reduced
from its resonance values. We also find that the position of the
$Q^*$ mode is almost frequency independent, with a maximum
intensity at $\omega \approx 54$ meV, in agreement with the
experimental observations (see Fig.~2a of Ref.~\cite{pai}).

In what follows, we discuss the physical origin of our results. We
begin by reviewing the emergence of the $Q$ mode.
Its origin lies in a feedback effect on the
spin excitation spectrum (i.e., on $\chi_0$) due to the opening of
the SC gap. This feedback effect is universal in that it only
requires that the FS possesses points, ${\bf k}$ and ${\bf
k^\prime}$, that can be connected by ${\bf Q}$ (so-called ``hot
spots") and that ${\rm sgn}(\Delta_{\bf k})=-{\rm sgn}(\Delta_{\bf
k^\prime})$. To demonstrate the universality of the resonance
mode, we rewrite the denominator of Eq.~(\ref{fullchi}) as
\begin{equation}
\xi^{-2}_{\bf Q}-g({\bf Q}) \Delta
\chi_0 ({\bf Q}, \Omega)
\end{equation}
where $\xi^{-2}_{\bf Q} = 1-g({\bf Q}) \chi^{ns}_0({\bf Q},0)$ and
$\Delta \chi_0 ({\bf Q}, \Omega) = \chi_0 ({\bf Q}, \Omega) -
\chi^{ns}_0 ({\bf Q}, 0)$. The main contribution to the static
normal state susceptibility $\chi^{ns}_0 ({\bf Q}, 0)$ comes from
high internal frequencies in the fermionic bubble. This implies
that $\chi^{ns}_0$ and thus $\xi^{-2}_{\bf Q}$ are
non-universal quantities that depend on the details of the
band structure. However, the
appearance of the resonance mode only requires $\xi^{-2}_{\bf
Q}>0$, which is satisfied in the paramagnetic state. In contrast,
$\Delta \chi_0 ({\bf Q}, \Omega)$ is {\it universal} since its
main contribution comes from small internal frequencies of order
$\Delta$. $\Delta \chi_0$ can therefore be evaluated by simply
linearizing the fermionic dispersion near ${\bf k}$ and ${\bf
k^\prime}$. The integration over momentum~\cite{comm2} yields at
$T=0$
\begin{eqnarray}
\Delta \chi_0 ({\bf Q}, \Omega) &=& -i \frac{\gamma^{NS}_{\bf
Q}}{16} \sum_{\bf \{
k,k^\prime\}} \int_{-\infty}^\infty d\omega \nonumber \\
& \times & \left(1 -\frac{\omega_+ \omega_- + \Delta_k
\Delta_{k^\prime}}{\sqrt{\omega^2_{+} - \Delta^2_k}
\sqrt{\omega^2_- - \Delta^2_{k^\prime}}}\right) \label{n1}
\end{eqnarray}
where $\omega_{\pm} = \omega \pm \Omega/2$, and the summation runs
over all pairs of FS points ${\bf k}$ and ${\bf k^\prime}$
separated by ${\bf Q}$. There are eight fermionic scattering
processes from ${\bf k}$ to ${\bf k}^\prime$ in the first zone. The
momenta ${\bf k}$ involved are those in which the boundary of the
magnetic zone (defined by $\cos k_x =-\cos k_y$) crosses the FS. Two of
these scattering processes are direct with ${\bf k-k}^\prime= {\bf
Q}$, two involve umklapp scattering with ${\bf k-k}^\prime= {\bf
Q}-(2 \pi,2\pi)$, and four involve umklapp scattering with ${\bf
k-k}^\prime= {\bf Q}-(2\pi,0)$ and ${\bf k-k}^\prime= {\bf
Q}-(0,2\pi)$. A pair of FS points that are connected by direct
scattering via ${\bf Q}$ is shown in Fig.~\ref{FS}(a) (${\bf Q}$
is represented by a dashed arrow).

In the normal state, all eight processes equally contribute to
$\Delta \chi_0 ({\bf Q}, \Omega)$, and one obtains from
Eq.~(\ref{n1}) $\Delta \chi_0 ({\bf Q}, \Omega) = - i
\gamma^{NS}_{\bf Q} \Omega$ which identifies $\gamma^{NS}_{\bf Q}$
with the Landau damping rate \cite{comm3}. In the superconducting
state, $\Delta \chi_0$ becomes a complex function. According to
Eq.~(\ref{n1}), its imaginary part vanishes below a critical
frequency $\Omega_c({\bf Q}) = |\Delta_{\bf k}| + |\Delta_{{\bf
k}^\prime}|$, which is the same for all eight scattering channels.
The $d_{x^2-y^2}$ symmetry of the SC gap implies $\Delta_{\bf
k}=-\Delta_{{\bf k}^\prime}$, resulting in a discontinuous jump of
Im$\chi_0$ at $\Omega_c({\bf Q})$ from zero to $\pi
\gamma^{NS}_{\bf Q} |\Delta_{\bf k}|$~\cite{excit}.
Simultaneously, Re$\Delta \chi_0 ({\bf Q}, \Omega)>0$ is non-zero,
diverges logarithmically at $\Omega_c({\bf Q})$, and scales as
$\Omega^2$ at small frequencies. Re$\Delta \chi_0$ therefore
varies between $0$ at $\Omega=0$ and $\infty$ at
$\Omega=\Omega_c$. Since $\xi^{-2}_{\bf Q}>0$, one finds that for
any positive $g_{\bf Q}$, $\chi({\bf Q}, \Omega)$
[Eq.~(\ref{fullchi})] acquires a pole at a frequency $\Omega_{res}
< \Omega_c({\bf Q})$ where Re$ \Delta \chi_0 ({\bf Q},
\Omega_{res})=\xi^{-2}_{\bf Q}/g_{\bf Q}$ and Im$\Delta \chi_0
({\bf Q} , \Omega_{res})=0$. Im$\chi$ thus exhibits a
$\delta-$function at $\Omega_{res}$, representing a spin exciton
below the particle-hole (p-h) continuum.

We next consider the dispersion of the exciton. For ${\bf q} \neq
{\bf Q}$, the degeneracy of the scattering channels is lifted. In
particular, for momenta along ${\bf q} = \eta (\pi,\pi)$, one now
has three different critical frequencies, $\Omega_c^{(i)}({\bf
q})$ ($i=1,2,3$) \cite{excit}.  $\Omega_c^{(1)}$ is the critical
frequency associated with direct scattering, $\Omega_c^{(2)}$ with
umklapp scattering involving ${\bf q}-(2\pi,0)$ and ${\bf
q}-(0,2\pi)$, and $\Omega_c^{(3)}$ with umklapp scattering by
${\bf q}-(2 \pi, 2\pi)$. The scattering momenta for ${\bf q}$
close to ${\bf Q}_0$ are shown in Fig.~\ref{FS}(a), and
$\Omega_c^{(i)}({\bf q})$ for ${\bf q}=\eta (\pi,\pi)$ are
presented in Fig.~\ref{FS}(b). For all of these scattering
processes, $\Delta_{\bf k}$ and $\Delta_{{\bf k}^\prime}$ still
have opposite signs \cite{sign}. As a result, Im$\Delta \chi_0$
exhibits three discontinuous jumps at $\Omega_c^{(i)}({\bf q})$,
and Re$\Delta \chi_0$ diverges logarithmically at
$\Omega_c^{(i)}({\bf q})$~\cite{excit}. However, Im$\Delta \chi_0$
is zero only below the smallest $\Omega_c^{(i)}({\bf q})$, and
hence a true resonance is only possible below the smallest
critical frequency. The splitting of the critical frequencies for
${\bf q} \not = {\bf Q}$ was also discussed in Ref.~\cite{Morr00}.

It follows from Fig.~\ref{FS}(b) that for $0.8 < \eta < 1.2$, the
smallest critical frequency, $\Omega_c^{(1)}$, corresponds to
direct (i.e., non-umklapp) scattering. $\Omega_c^{(1)}$ decreases
away from $\eta =1$ and eventually vanishes at $\eta = 0.8$, when
direct scattering occurs between nodal points at the Fermi
surface~\cite{excit}. Since the exciton is necessary located below
$\Omega_c^{(1)}$, its frequency also decreases and eventually
vanishes at  $\eta = 0.8$. Moreover, upon approaching $\eta =0.8$
the jump in Im$\Delta \chi_0 ({\bf q}, \Omega_c^{(1)})$ decreases
as ${\bf k}$ and ${\bf k}^\prime$ approach the nodal points.
Accordingly, the resonance frequency $\Omega_{res}$ moves closer
to $\Omega_c^{(1)}$, and the intensity of the resonance
decreases~\cite{excit}.

At ${\bf Q}_0 = 0.8 (\pi,\pi)$,  Im$\chi_0 ({\bf Q}_0, \Omega)$ is
non-zero for $\Omega>0$, and one finds Im$\chi({\bf Q}_0, \Omega)
= \gamma^{SC}_{{\bf Q}_0} \Omega$ at small frequencies, where
$\gamma^{SC}_{{\bf Q}_0} =\gamma^{NS}_{\bf Q} \frac{1}{8}
\frac{\pi v_F}{4 v_\Delta}$ and $v_\Delta$ is the gap velocity at
the nodal points. The factor $1/8$ arises since only a single
(direct) scattering channel contributes to $\gamma^{SC}_{{\bf
Q}_0}$, while eight channels contribute to $\gamma^{NS}_{{\bf
Q}}$. However, since the Fermi velocities at the nodal points are
antiparallel, $\gamma^{SC}_{{\bf Q}_0}$ depends on $v_F$ only
through the ratio $v_F/v_\Delta \sim 20$ which compensates the
small prefactor \cite{comm4}. As a result, $\gamma^{SC}_{{\bf
Q}_0}$ is comparable to $\gamma^{NS}_{{\bf Q}}$, thus giving rise
to a weak and featureless frequency dependence of Im$\chi({\bf
Q}_0,\Omega)$ similar to that of Im$\chi_0 ({\bf Q}, \Omega)$ in
the normal state. The vanishing of the gap in the p-h continuum at
${\bf Q}_0$ together with the large value of $\gamma^{SC}_{{\bf
Q}_0}$ explains the experimental observation of a ``silent band"
in Ref.~\cite{pai} (the position of which is indicated by the
shaded area in Fig.~\ref{exper}).

For momenta ${\bf q} < {\bf Q}_0$, (i.e., $\eta<0.8$) the nodal
points cannot be connected, and a {\it direct} gap opens for
excitations into the p-h continuum. This gap is independent of the
SC gap, and given by $\Omega^d_c={\bf v}_F \cdot ({\bf Q}_0 -{\bf
q})$ (see dashed-dotted line in Fig.~\ref{FS}(b)). Due to a large
$|{\bf v}_F|$, $\Omega^d_c$ becomes equal to $\Omega_c^{(2)}$
close to ${\bf Q}_0$ at ${\bf q}_d= \eta_d {\bf Q}$ with
$\eta_d=0.773$ (i.e., ${\bf Q}_0-{\bf q}_d= 0.027 {\bf Q}$ for the
dispersion of Eq.~(\ref{Disp})). For $\eta_d < \eta <0.8$, one
finds that Im$\chi_0$ vanishes below $\Omega^d_c$, and Im$\chi_0
\sim \sqrt{\Omega-\Omega^d_c}$ for $\Omega>\Omega^d_c$. Hence,
Re$\chi_0$ does not diverge at $\Omega^d_c$, and no resonance peak
exists in this region, extending the silent band. However,
Re$\chi_0$ possesses a logarithmic divergence at $\Omega_c^{(2)}$,
and hence it satisfies the resonance condition Re$ \Delta \chi_0
({\bf q}, {\bar \Omega}_{res})=\xi^{-2}_{\bf q}/g_{\bf q}$, at
some frequency ${\bar \Omega}_{res}$ below $\Omega_c^{(2)}$. Once
$\Omega^d_c$ crosses ${\bar \Omega}_{res}$, the damping at ${\bar
\Omega}_{res}$ vanishes and a true pole in Im$\chi$ occurs,
leading to the appearance of the $Q^*$ mode (open circles in
Fig.~\ref{FS}(b)). As one moves further away from ${\bf Q}_0$, one
finds that the $Q^*$ resonance is rapidly suppressed. This
suppression arises from the rapid decrease of $\Omega_c^{(2)}$ as
well as the decrease of the bare static spin susceptibility,
$\chi_0({\bf q},0)$. This behavior of $\chi_0 ({\bf q}, 0)$ is
similar to that in a system of free fermions with a circular Fermi
surface, if we identify ${\bf Q}_0$ with $2 {\bf p}_F$. Both
effects lead to a rapid shift of the $Q^*$ mode towards the edge
of the p-h continuum, and to a decrease in its intensity. As a
result, this resonance is only visible near ${\bf Q}_0$.

\begin{figure}[h]
\epsfig{file=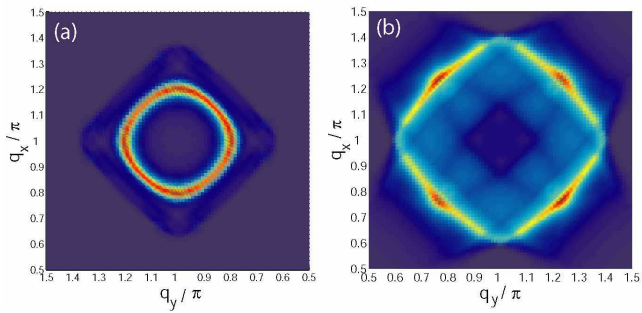,width=8.5cm} \caption{(color) Intensity plot
of Im$\chi$, Eq.~(\ref{fullchi}), as a function of momentum (blue
(red) implies a small (large) value of Im$\chi$) at: (a) $33$ meV
($Q$ mode), and (b) $57$ meV ($Q^*$ mode). The $Q$ mode forms a
distorted ring in momentum space, with intensity maxima along
$(\pi,q)$. In contrast, the intensity of the $Q^*$ mode is largest
along the zone diagonal. In (b), the periodic structure of Im$\chi$
around ${\bf Q}$ reflects the momentum dependence of the p-h
continuum at this energy.} \label{fig4}
\end{figure}

The $Q$ and $Q^*$ modes are not only separated in frequency, as
discussed above, but their intensity maxima are also located in
different parts of the MBZ; this represents a major qualitative
distinction between the two modes. In Fig.~\ref{fig4} we present
intensity plots of Im$\chi$ as a function of momentum for $\Omega
> \Omega_{res}({\bf Q})$, probing the $Q^*$ mode
[Fig.~\ref{fig4}(b)], and for $\Omega < \Omega_{res}({\bf Q})$,
probing the $Q$ mode [Fig.~\ref{fig4}(a)]. The difference is
striking. While the intensity of the $Q$ mode is largest along
${\bf q}= (\pi, \eta \pi)$ and ${\bf q}=(\eta  \pi, \pi)$, the
$Q^*$ mode has its largest intensity along the diagonal direction,
i.e., along ${\bf q}=\eta(\pi,\pi)$ and ${\bf q}=[(2-\eta) \pi,
\eta \pi]$.
This rotation of the intensity pattern by $45^\circ$ reflects the
qualitative difference in the origin of the two modes. The
intensity of the $Q$ mode is at a maximum along ${\bf q}=( \pi,
\eta \pi)$ and ${\bf q}=(\eta \pi, \pi)$, since in this case the
fermions that are scattered by ${\bf q}$ are located farther from
the nodes than for diagonal scattering. In contrast, the $Q^*$
mode arises from the rapid opening of a gap in the p-h continuum
below ${\bf Q}_0$, which is most pronounced along the diagonal
directions of the zone. The rotation of the intensity pattern by
$45^\circ$, which is the most significant distinction between the
$Q$ and $Q^*$ modes, was observed in recent INS experiments
\cite{Hay04,mike}.

In summary, we showed that a new resonant magnetic excitation at
incommensurate momenta, observed recently by inelastic neutron
scattering experiments on YBa$_2$Cu$_3$O$_{6.85}$ \cite{pai} and
YBa$_2$Cu$_3$O$_{6.6}$ \cite{Hay04}, is a spin exciton arising
from umklapp scattering. Its location in the zone and its
frequency are determined by the momentum dependence of the p-h
continuum. It is confined to a small region in ${\bf q}$ just
below ${\bf Q}_0$, and is separated from the $Q$ resonance by a
``silent band" where Im$\chi$ is strongly suppressed. We also
found that the intensity maxima of the two modes are rotated by
$45^\circ$ relative to each other.

We thank Ph.~Bourges, B.~Keimer, and H.~Mook for discussions
concerning their data. I.E. is thankful to INTAS (No. 01-0654) and
RSP Superconductivity Grant No. 98014-3 for the support. D.K.M.
acknowledges support from the Alexander von Humboldt foundation.
A.C. acknowledges support from NSF DMR 0240238 and SFB 290. M.R.N.
is supported by the US Dept. of Energy, Office of Science, under
Contract No. W-31-109-ENG-38.  D.K.M. and A.C. are grateful for
the hospitality of the Freie Universit\"{a}t Berlin.  D.K.M.,
A.C., and M.R.N. also acknowledge support from the Aspen Center
for Physics where this work was completed.\\

*on leave from the Physics Dept., Kazan State University, 420008,
Kazan, Russia.


\begin{thebibliography}{99}
\bibitem{com5} For the numerical calculation of $\chi_0$,
we employed $\delta=2$ meV in the analytic continuation of the
Greens functions $i \omega_n \rightarrow \omega+i\delta$.
\bibitem{comm2} This expression is
valid even for interacting fermions as long as the fermionic
self-energy does not depend on the momentum transverse to the
Fermi surface. We also assume for simplicity that $\Delta$ is
independent of frequency.
\bibitem{comm3} Re$\Delta \chi_0 \not = 0$
only if one includes the curvature of the dispersion in the normal
state.
\bibitem{sign} For small enough $q$, this sign change no longer
occurs.  In particular, for $\Omega_c^{(2)}$, there is no sign change when $q$
is less than the ``kink" in Fig.~3(b).
\bibitem{mike}  The underdoped INS data reveal a pattern in momentum
space which is more spot like than the optimal doped data, both
for the $Q$ and $Q^*$ modes.  This effect can be reproduced in the
RPA calculations if the Fermi surface is flattened near the node.
\end{thebibliography}
\end{document}